# Janus β-PdXY (X/Y = S, Se, Te) Materials with high Anisotropic Thermoelectric Performance

Mukesh Jakhar[1], Raman Sharma[2] and Ashok Kumar[1*]

[1]*Department of Physics, School of Basic Sciences, Central University of Punjab, Bathinda, 151401, India*

[2]*Department of Physics, Himachal Pradesh University, Shimla, 171005, India*

(January 28, 2023)

*Corresponding Author:

ashokphy@cup.edu.in (Ashok Kumar)

**Abstract**

Two-dimensional (2D) materials have garnered considerable attention as an emerging thermoelectric (TE) material owing to their unique density of state (DOS) near the Fermi level. We investigate the TE performance of Janus β-PdXY (X/Y=S, Se, Te) monolayer materials as a function of carrier concentration and mid-temperature range (300 to 800 K) by combining density functional theory (DFT) and semi-classical Boltzmann transport theory. The phonon dispersion spectra and AIMD simulations confirm their thermal and dynamical stability. The transport calculation results reveal the highly anisotropic TE performance for both n and p-type Janus β-PdXY monolayers. Meanwhile, the coexistence of low phonon group velocity and converged scattering rate leads to lower lattice thermal conductivity ($K_l$) of 0.80 W/m K, 0.94 W/m K, and 0.77 W/m K along y-direction for these Janus materials. While the high TE power factor is attributed to the high Seebeck coefficient (S) and electrical conductivity, which is due to the degenerate top valance bands of these Janus monolayers. The combination of lower $K_l$ and high-power factor at 300K (800 K) leads to an optimal figure of merit (ZT) as 0.68 (2.21), 0.86 (4.09) and 0.68 (3.63) for p-type Janus PdSSe, PdSeTe and PdSTe monolayers. To capture rational electron transport properties, the effects of acoustic phonon scattering ($\tau_{ac}$), impurity scattering ($\tau_{imp}$), and polarized phonon scattering ($\tau_{polar}$) are included in the temperature-dependent electron relaxation time. These findings indicated that the Janus β-PdXY monolayers are promising candidates for TE conversion devices.

## 1. Introduction

The environmental-friendly energy materials, such as thermoelectric (TE) materials, can convert ubiquitous waste heat to electricity through thermoelectric technologies, thus, providing a promising solution for energy shortages and the environmental pollution crisis.[1-4] Generally, the performance of TE materials as solid-state TE power generators is expressed by the dimensionless figure of merit (ZT) quantity.[5] The low conversion efficiency of traditional TE materials limits the development and commercialization of TE devices on a broad scale. The high power factor (PF), low conductivity, or both, simultaneously is an essential evaluating parameter for the optimal value of ZT to make the TE generators competitive with other alternative energy sources.[6, 7]

To date, much efforts have been put into improving TE conversion efficiency, which entails both increasing electron mobility and lowers thermal conductivity. The band engineering[8-10] and nanostructure engineering[11-13] are the most effective approaches to tackle the electrical and thermal transport properties for achieving an overall high ZT value. Anisotropic materials, such as SnSe, SnS[14], LaCuOSe,[15] $Mg_3Sb_2$ [16], $Sb_2Si_2Te_6$ [17], KBaBi [18] and many more, have recently been discovered to have characteristics that are very distinct in different directions, making them appealing for TE energy conversion. Moreover, the anisotropic two dimensional (2D) materials also piqued interest, as these provide extra TE benefits due to their low dimensionality.[19-21] Therefore, 2D-TE materials have stimulated the interest of several scientists due to the enhancement of electronic DOS at the Fermi level, which can enhance the Seebeck coefficient (S) according to the Mott equation.[22] Notably, the high S can lead to the high performance of TE materials.[23]

Janus 2D materials, which are new variants of 2D materials, have gained much research attention because of their distinct properties. Due to mirror asymmetry in crystal structure, they exhibit unique properties as compared to pristine 2D materials.[24, 25] From the family of 2D transition metal dichalcogenides (TMDs), the Janus MoSSe monolayer was experimentally synthesized from their pristine $MoS_2$ monolayers as well from $MoSe_2$ monolayers.[26, 27] Janus monolayers show great potential in variety of applications such as optoelectronics[28, 29], photocatalysts[30-33], gas sensing[34], spintronics[35, 36], valleytronics[37] and photovoltaic.[38] Also, Gu and Yang[39] demonstrated that by reducing frequency gap, the lattice thermal conductivity of the materials could be significantly reduced. It can also be decreased by altering the TMD stoichiometric ratio.[40-42]

Subsequently, the Group-10 based metal (Pd, Pt) dichalcogenides materials exhibit potential in terms of thermoelectric properties, e.g., ZT value of pentagonal $PdTe_2$,[43] $PtTe_2$,[44] and hexagonal Janus MXY (M = Pd, Pt; X, Y = S, Se, Te)[45] monolayer is calculated to be 2.42, 2.60 and 0.58-2.64, respectively, using first principles theory. The possibility of the existence in different phases, the thermoelectric properties of other phases of Group 10 based metal (Pd, Pt) dichalcogenides material need extensive investigation. Therefore, we systematically explore the thermoelectric performance of the Janus materials of the recently reported stable β-phase of $PdX_2$ (X=S, Se, Te)[46, 47] monolayers.

In this study, we conduct extensive research of the phononic and thermal transport properties to estimate TE performance for 2D Janus β-PdXY (X/Y=S, Se, Te) monolayers by utilizing Boltzmann transport theory. Here the TE properties of these Janus monolayers are explored as a function of carrier concentration and a temperature range from 300 to 800 K for both p- and n-type doping along different direction. We found that all these Janus monolayers exhibit anisotropic power factor and thermal conductivity. Meanwhile, at 300 K, the low lattice thermal conductivity (high ZT value) of Janus PdSSe, PdSeTe and PdSTe monolayers were found as low (high) as 0.80 W/m K (0.76), 0.94 W/m K (0.86) and 0.77 W/m K (0.71), respectively, that predicts Janus β-PdXY (X/Y=S, Se, Te) monolayers to be promising TE materials.

## 2. Computational Details

Here, the first principles simulation are performed for the 2D Janus β-PdXY (X/Y=S, Se, Te) monolayers within Density functional theory (DFT) using VASP Simulation Package [48, 49]. The electron exchange-correlation functional was implemented by the GGA (generalized gradient approximation)[50] parameterized of Perdew−Burke−Ernzerhof (PBE) scheme[51]. Meanwhile, the Heyd−Scuseria− Ernzerhof (HSE06)[52] level of theory is also employed to simulate the band structure of Janus monolayers. The kinetic energy cut-off of 600 eV is set for the plane-wave basis set[51]. The ionic relaxation is conducted until all atomic forces and energy are less than 0.001 eV/Å and $10^{-8}$ eV, respectively. A vacuum of 20 Å along Z-direction is adopted to avoid mutual interactions between adjacent layers. The Monkhorst pack of 24 × 24 ×1 k mesh was used for the Brillouin zone sampling[53]. Moreover, to investigate the thermodynamic stability at 300 and 500 K, the ab initio molecular dynamics computation was performed with a time step of 1 fs.

The thermal properties were estimated using ShengBTE code[54]. The phonon dispersion curves were obtained by DFPT (density functional perturbation theory)[55] method with PHONONPY packages[56]. To ensure the convergence, the 2$^{nd}$ (harmonic) and 3$^{rd}$ (anharmonic) order interaction force constants (IFCs) were obtained by using large supercells of $6 \times 6 \times 1$ and $4 \times 4 \times 1$ with k-mesh of $4 \times 4 \times 1$ and $5 \times 5 \times 1$, respectively. It is well known that the acoustic out-of-plane flexural (ZA) phonon dispersion plays a central role in determining many properties such as phonon thermal transport[57], electron-phonon coupling[58], and thermodynamic stability.[59] Thus, for a more accurate and efficient thermal conductivity estimation of Janus β-PdXY (X/Y=S, Se, Te) monolayers, the obtained 2$^{nd}$ harmonic IFCs (here called as "raw") were considered to enforce the rotational invariance condition as a post processing step to get "corrected" IFCs by using the HiPhive package[60] for all three Janus monolayers.

To get an accurate 3$^{rd}$ IFC (Interaction Force Constants), convergence tests reveals an interaction ranges up to 7.10 Å, 7.52 Å and 7.33 Å distance with 13$^{th}$ nearest neighbor atoms for all the Janus PdSeS, PdSeTe and PdSTe monolayers. Meanwhile, for computing the lattice thermal conductivity, a high dense grid of $36 \times 36 \times 1$ has been used for sampling to ensure convergence. The thermal conductivity at all temperatures is calculated by iterative solution of Boltzmann transport equation. Note that the phonon-isotope scattering is taken into consideration in our calculations. Moreover, the convergence test of nearest neighbors, Q-grid and Gaussian smearing for calculating the lattice thermal conductivity are available in ESI.

Further, the electronic transport properties were carried out using BoltzTrap2 code[61, 62] based on the BTE (Boltzmann transport equation) within the rigid band approximation. The denser 10000 k-points in irreducible Brillouin Zone were used to simulate the electronic transport properties. In the electrical and thermal conductivity results, the (z/d) correction is carried out, where the z is the unit cell length along the Z-axis and d is the effective thickness of the 2D monolayer. Here the effective thickness (thickness of monolayer along Z-axis + van der Waals radius of surface atoms)[63, 64] of Janus PdSSe, PdSeTe and PdSTe monolayers are estimated to be 3.63 Å, 2.82 Å and 3.24 Å, respectively. The results of the thermoelectric figure of merit ZT are independent of this correction because we use the same effective thickness for both electrical and phonon transport property calculations.

## 3. Results and Discussion

### 3.1 Geometric and Electronic Structure

The crystal structure (top and side view) of Janus β-PdXY (X/Y=S, Se, Te) is shown in Fig. 1. Unlike hexagonal T-phase (four-membered rings) and octahedral pentagonal (five-membered rings) forms[65], the Janus β-PdXY is composed of six and four-membered ring structures parallel to the x-axis as a helical chain similar to its pristine β-PdX$_2$ monolayers.[46, 47] Each Pd and X/Y atom is bonded with four X/Y atoms and two adjacent Y/X atoms to form the Pd−X/Y and X/Y-X/Y bond. The optimized parameters of Janus PdSSe, PdSeTe and PdSTe monolayer are listed in Table 1. The lattice parameters of Janus monolayers lies in between their parent β-PdS$_2$, β-PdSe$_2$, β-PdTe$_2$ monolayers.[46, 47]

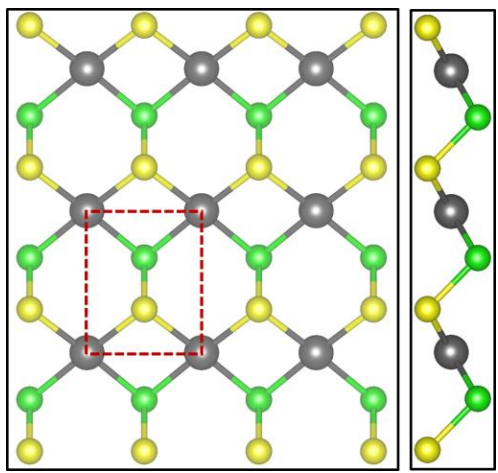

**Fig. 1** The structure (top view and side view) of Janus β-PdXY (X/Y =S, Se, Te) monolayers. The red dotted line represents the unit cell for the Janus monolayers. The black, orange and green spheres represent Pd, X, and Y atoms, respectively.

The thermal stability of Janus β-PdXY (X/Y=S,Se,Te) monolayer is also carried out via AIMD simulations at room temperature (300 K) and high temperature of 500 K using relatively larger 4 × 4 × 1 supercell. The small thermal fluctuation in temperature and total energy as a function of the simulation time steps and optimized ground-state structures suggest that Janus β-PdXY (X/Y=S,Se,Te) monolayers are thermally stable (Fig. S1-3 (a) ESI) at room temperature (300 K), thereby showing promises to the future synthesis of these β phases Janus monolayers.[66] It is noted that the temperature fluctuation increases with function of time steps at higher temperature (500K) as shown in Fig. S1-3 (b) ESI which results in small distortion in the structural parameters at the end of 5000 fs seconds (inset in Fig. S1-3 ESI), that leads to less stable structures at higher temperature.

**Table 1:** The optimized parameters: lattice constants (Å), bond length (Å), and band gap (eV) for the Janus β-PdXY (X/Y =S, Se, Te) monolayers.

| Janus | Lattice Constants (Å) | | Bond Length, (X/Y = S, Se, Te) (Å) | | | Bang Gap (eV) | |
|---|---|---|---|---|---|---|---|
| | a | b | Pd-X | Pd-Y | X-Y | PBE | HSE06 |
| PdSSe | 3.57 | 4.46 | 2.37 | 2.46 | 2.28 | 1.10 | 2.04 |
| PdSeTe | 3.80 | 4.68 | 2.52 | 2.60 | 2.59 | 0.82 | 1.67 |
| PdSTe | 3.71 | 4.59 | 2.41 | 2.59 | 2.44 | 0.87 | 1.71 |

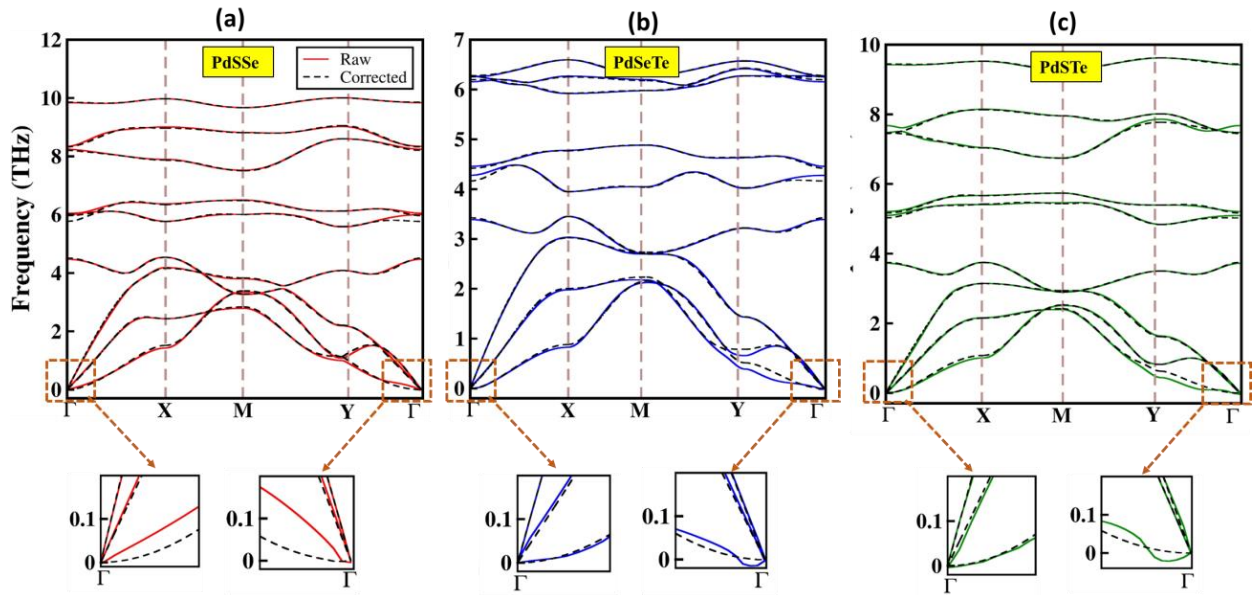

**Fig. 2** The calculated phonon dispersion curves for Janus (a) PdSSe (b) PdSeTe (c) PdSTe monolayers with zoom-in near Γ point represent raw and corrected acoustic phonons modes.

Next, the dynamic stability of the Janus β-PdXY (X/Y=S, Se, Te) monolayer is assessed by examining the phonon dispersion spectra along the high symmetry direction of Brillouin zones. All three Janus PdSeS, PdSeTe and PdSTe has three atoms in each unit cell shown in the inset of Fig. 1, so possesses three acoustic (a transverse acoustic mode (TA), a longitudinal acoustic mode (LA), and a flexural acoustic mode (ZA)) and six optical phonon modes corresponding to their one metal (Pd) atom and two chalcogen (X/Y = S, Se, Te) atoms in the unit cell. Different from three linear acoustic phonon modes in 3D, there is an unique and interesting out-of-plane (ZA) mode in 2D materials[67, 68]. Note that the ZA mode is an out-of-plane transverse acoustic mode which is quadratic near Γ point similar to other 2D materials like graphene[69], phosphorene[70], and stanene [71] (Fig. 2 (a-c)). The highest vibration frequencies of the acoustic (optical) branches are about 4.18 (9.97) THz, 3.03 (6.59) THz and 3.10 (9.52) THz for the Janus PdSSe, PdSeTe and PdSTe monolayer, respectively. According to previous studies, the low-frequency acoustic phonon modes significantly affect thermal conductivity.[70, 72] The Janus PdSeTe and PdSTe monolayer have lower acoustic phonon spectra region as compared to Janus PdSSe monolayer, which may lead to lower thermal conductivities for PdSeTe and PdSTe monolayer than the PdSSe monolayer.

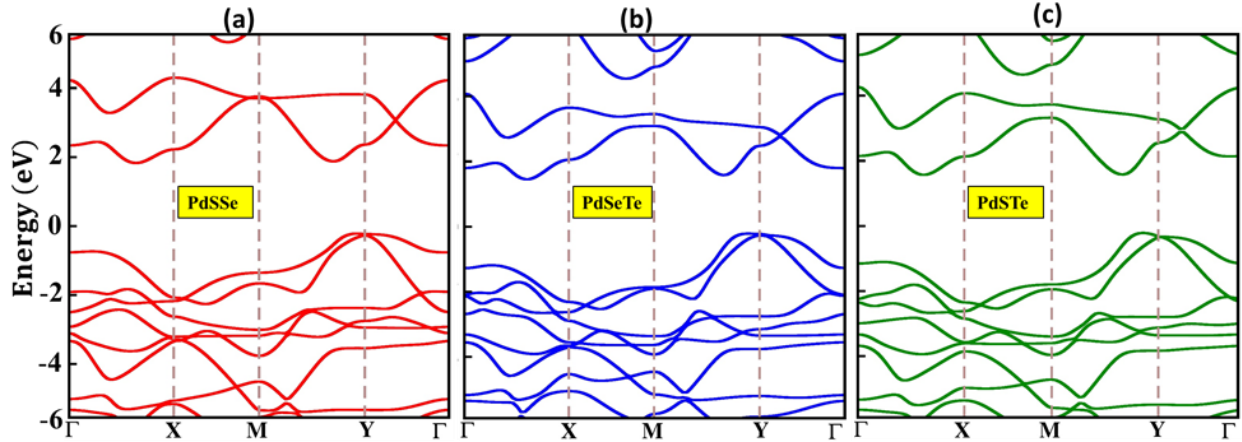

**Fig. 3** The computed electronic band structure at HSE06 level for Janus (a) PdSSe (b) PdSeTe (c) PdSTe monolayers.

Now, the electronic band structure of Janus β-PdXY (X/Y =S, Se, Te) monolayers is computed using both PBE and HSE06 levels of theories. These monolayers exhibit indirect band gap of 2.04 eV (1.10 eV), 1.67 eV (0.82 eV), and 1.71 eV (0.87 eV) for Janus PdSSe, PdSeTe and PdSTe monolayers, respectively, using HSE06 (PBE) level of theories (Fig. 3 (Fig. S4)). The obtained

HSE06 band gap for all the three Janus monolayers are in between that of their pristine $PdS_2$ (2.10 eV)[47], $PdSe_2$ (1.94 eV)[46] and $PdTe_2$ (1.29 eV)[47] monolayer. Density of states (DOS) calculations reveal that in the valence states near Fermi level, the chalcogens (S, Se, Te) atoms contribution are coequally dominated for Janus β-PdXY (X/Y =S, Se, Te) monolayers (Fig. S5). In contrast, the Palladium atom contribution is high near the Fermi level for all the three Janus β-PdXY (X/Y =S, Se, Te) monolayers for the conduction states. We also compute the effect of spin-orbit coupling (SOC) using PBE+SOC functional. The SOC effect on Janus PdSSe is negligible, whereas a slight band gap reduction was obtained for PdSeTe (0.17 eV) and PdSTe (0.14 eV) due to heavy Te atom (Fig. S4). Notably, towards the Fermi level, a few of the top valence bands are energetically degenerate (Fig. S4). These degenerate conduction bands may result in a higher Seebeck coefficient, and a higher power factor (PF) for p-type doping compared to n-type doping.

### 3.2 Thermal Transport Properties

The calculated lattice thermal conductivity ($K_l$) of stable Janus β-PdXY (X/Y=S, Se, Te) monolayer with temperature is plotted along x and y-direction in Fig. 4(a). We find that, as temperature increases, the $K_l$ of the monolayers gradually decreases which satisfy the temperature dependence relationship ($K_l \propto 1/T$), suggesting that the $K_l$ is dominated by anharmonic phonon−phonon interactions with temperature [73]. The estimated $K_l$ for the Janus PdSSe, PdSeTe, and PdSTe monolayer at 300 K are anisotropic and as low as 10.47 (0.80) W/m K, 5.86 (0.94) W/m K, and 4.43 (0.77) W/m K along x (y)-direction, respectively.

Such ultralow $K_l$ at 300 K of Janus PdSTe and PdSeTe is much lower than the pentagonal $PdS_2$ (4.34) (12.48) W/m K,[43] $PdSe_2$ 2.91(6.62) W/m K, [43] $PdTe_2$ 1.42 (5.90) W/m K [43] along x(y) direction as well as hexagonal Janus PdSSe (10.65 W/m K), and comparable to hexagonal Janus PdSTe and (5.45 W/m K) monolayer[45] due to the low-frequency phonons modes (acoustic frequencies) and smaller frequency gaps of Janus PdSTe and PdSeTe monolayers compared to their other phases. At 600 K the $K_l$ for Janus PdSTe (0.37 W/m K) monolayer along y direction is comparable to KAgSe (0.33 W/m K)[74], and δ-$Cu_2S$ (0.27 W/m K)[75] monolayers. These low thermal conductivities indicate that Janus β-PdXY (X/Y=S, Se, Te) monolayers could be excellent thermoelectric materials.

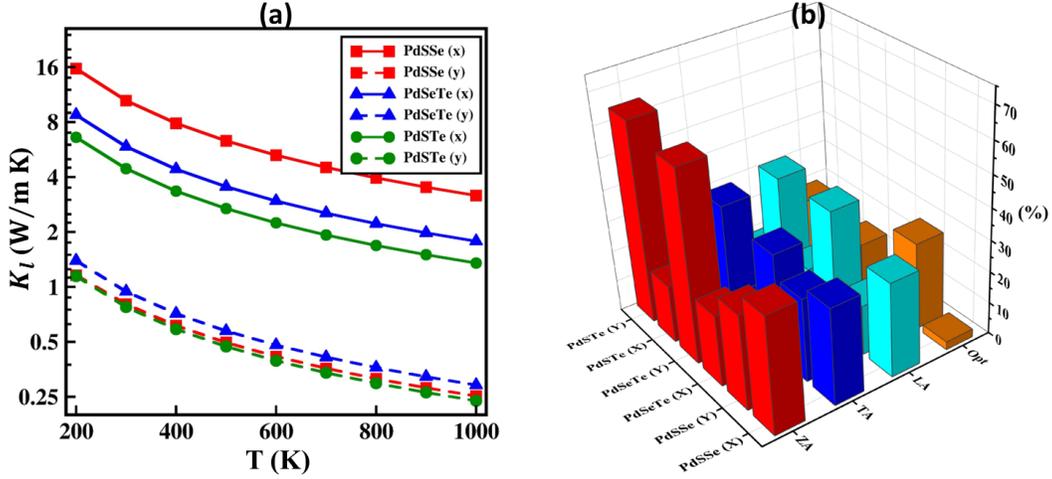

**Fig. 4** (a) The computed lattice thermal conductivity ($K_l$) and (b) their phonon mode contributions along x and y direction for Janus β-PdXY (X/Y=S, Se, Te) monolayers.

In order to get deeper insight into the ultralow thermal conductivities, we analyze the contribution of the phonon mode branches to the lattice thermal conductivity ($K_l$) of Janus β-PdXY (X/Y= S, Se, Te) monolayer as shown in the bar graph (Fig. 4(b)). The contribution to total thermal conductivity is mostly from the acoustic branch regardless of x- or y-direction for Janus β-PdXY (X/Y =S, Se, Te) monolayers. While optical branches for all three Janus monolayers have a negligible contribution to the $K_l$ along both x and y-direction. The out-of-plane ZA branch dominates among the acoustic branches for the Janus PdSSe, PdSeTe and PdSTe monolayers with 29.9 %, 60.9 %, and 64.3 % contribution along y-direction, respectively. While along the x-direction, like graphene[76], the longitudinal acoustic (LA) branch of acoustic mode shows a prominent contribution of 39.41 % and 38.45 % to total thermal conductivity for the Janus PdSeTe and PdSTe monolayers (Fig. 4(b)).

For deep-understanding the mechanism of low lattice thermal conductivity of Janus β-PdXY (X/Y=S, Se, Te), we plot the group velocity of phonon modes along the x and y-direction (Fig. 5). The greater group velocity along the x-direction leads to higher thermal conductivity as compared to y-direction. The maximum group velocities along the x-direction are contributed from the LA branches which are as high as 4.69 km/s, 3.73 km/s, and 3.57 km/s for the Janus PdSSe and PdSeTe and PdSTe monolayer, respectively (Fig. 4(b)). While for the y-direction, the maximum group velocity is 2.39 km/s and 3.07 km/s for PdSTe and PdSeTe monolayers (Fig. 5(b,c). At 300 K, the lowest group velocity along the x and y directions for PdSTe monolayers lead to the lowest thermal conductivity of 0.77 W/m K and 4.43 W/m K, respectively.

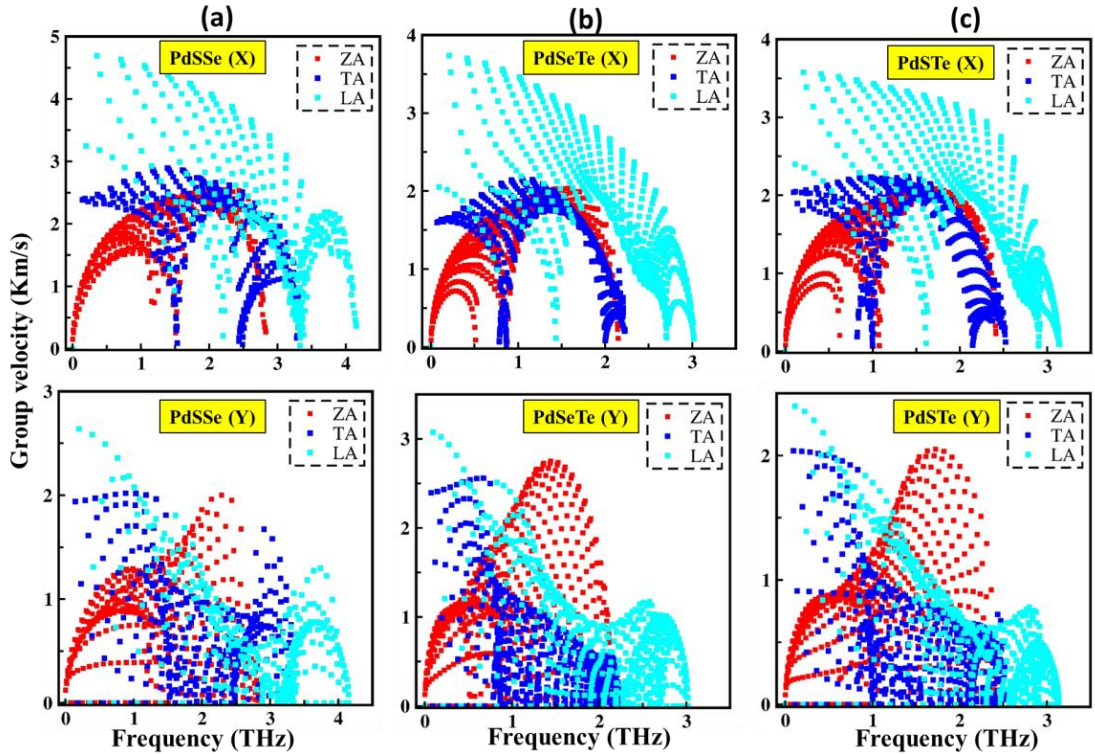

**Fig. 5** The computed acoustic phonon modes group velocity (Km/s) with frequency (THz) for Janus (a) PdSSe (b) PdSeTe (c) PdSTe monolayers along the x and y-direction.

In addition, the total converged phonon scattering rates is estimated at 300 K, as shown in Fig. 6. It shows that the three acoustic phonons (ZA, TA, LA) and six optical phonon branches are mainly occupied in the low-frequency and high-frequency portion, respectively, for all three Janus PdSSe, PdSeTe and PdSTe monolayers. However, the LA branch of acoustic mode overlaps with an optical mode that is commensurate with the frequency (THz) distribution of these three Janus materials' phonon spectra (Fig. 2). Besides the phonon scattering rates, the phonon phase space (P3) is presented in the inset of Fig. 6 for all three Janus PdSSe, PdSeTe and PdSTe monolayers. Meanwhile, as per $K_l \propto 1/P3$ relationship, the large total phase space leads to a lower value of $K_l$.[77] From the inset of Fig. 6(b), it is observed that the largest phonon space-time is $2.17 \times 10^{-5}$ eV$^{-1}$ for the PdSTe that leads to its lower value of $K_l$. Thus, low phonon group velocity and space time leads to high inhibit phonon transport with ultra-low $K_l$ for Janus β-PdXY (X/Y=S, Se, Te) monolayers.

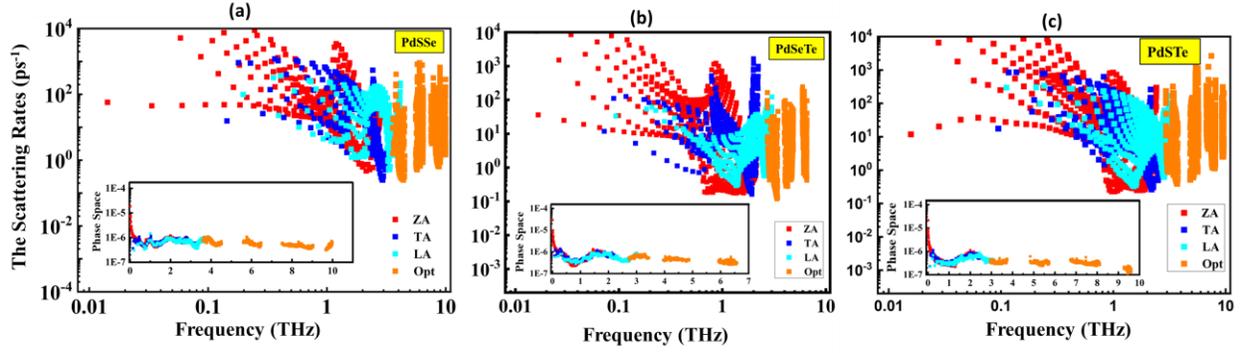

**Fig. 6** The converged phonon scattering rates (ps$^{-1}$) and phase space time (inset) with frequency (THz) for Janus (a) PdSSe (b) PdSeTe (c) PdSTe monolayers.

### 3.3 Carrier Mobility and Relaxation Time

The carrier mobility (μ) and electronic relaxation time ($\tau_e$) effectively impact the TE performance of materials. The Deformation theory (DP) of Bardeen−Shockley was used to estimate the carrier mobility including the acoustic phonon by the expression[14, 78]: $\mu_{2D} = \frac{2e\hbar^3 C_{2D}}{3K_B T m^* m_a^* E_i^2}$ where m*, $C_{2D}$ and $E_i$ are the effective mass, elastic modulus and deformation potential, respectively. According to the DP, m* and $C_{2D}$ describe the band curve and in-plane rigidity for the 2D system. On the other hand, $E_i$ ($\frac{\delta E_{edge}}{\delta \epsilon}$) is the first derivative of energy versus strain, denoting the shift of band edges under strain. Besides, ℏ is the reduced Planck constant, $k_B$ is Boltzmann constant, and T is the temperature. Note that here $m_a^*$ is the average effective mass given by $\sqrt{m_x^* m_y^*}$.

The computed highly anisotropic carrier mobility of p-type carrier for Janus PdSSe, PdSeTe and PdSTe monolayers are 106.02 (160.12) $cm^2V^{-1}S^{-1}$, 54.46 (863.86) $cm^2V^{-1}S^{-1}$ and 1424.79 (3618) $cm^2V^{-1}S^{-1}$ along x(y) direction, respectively. The high p-type carrier mobility along y-direction for Janus β-PdXY (X/Y=S, Se, Te) monolayers is due to the small value of the hole effective mass and deformation potential as comparable to n-type carrier (Table S1). Notably, the curvature of the parabola is inversely proportional to the effective mass, hence, strongly curved parabola will correspond to a low effective mass. Furthermore, the energy level is less affected by the variation of lattice potential field caused by the deformation of the valence band of Janus PdSTe monolayer along both x and y direction. Thus, it can be concluded that the weaker carriers-acoustic phonon coupling leads to smaller deformation potential for holes (Table S1). Thus the estimated p-type carrier mobility of Janus PdSTe monolayer are moderately large and much higher than that of their pristine monolayer such as β-PdS$_2$ (1069 cm$^2$V$^{-1}$S$^{-1}$)[47], PdTe$_2$ (1258 cm$^2$V$^{-1}$S$^{-1}$)[47] and also from Pentagonal PdS$_2$ (1664.15 cm$^2$V$^{-1}$S$^{-1}$) and PdSe$_2$ (3997 cm$^2$V$^{-1}$S$^{-1}$)[43] monolayers.

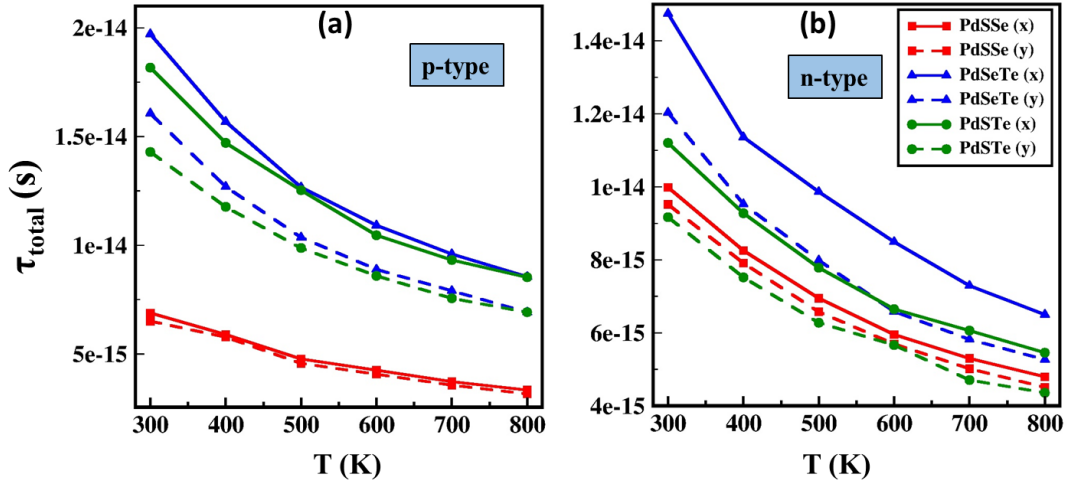

**Fig. 7** The calculated total electronic relaxation time ($\tau_{total}$) for (a) p-type and n-type Janus β-PdXY (X/Y =S, Se, Te) monolayers along the x and y-direction.

In addition, the temperature-dependent total relaxation time ($\tau_{tot}$) with the combination of acoustic phonon scattering ($\tau_{Acp}$), impurity scattering ($\tau_{Imp}$), and polarized phonon scattering ($\tau_{Pol}$) are estimated through the expression:

$$\frac{1}{\tau_{tot}} = \frac{1}{\tau_{Acp}} + \frac{1}{\tau_{Imp}} + \frac{1}{\tau_{Pol}} \qquad (1)$$

Among them, $\tau_{Acp}$ can be defined according to DP theory as:

$$\tau_{Acp} = \frac{\mu m_a^*}{e} \qquad (2)$$

While, the impurity scattering rate ($\tau_{Imp}$) and polarized phonon scattering ($\tau_{Polar}$) are calculated using a PYTHON code[79, 80]. The $\tau_{Imp}$ was expressed from the Brooks–Herring formula as:

$$\tau_{Imp} = \frac{\sqrt{2m^*}(4\pi\epsilon_0\epsilon_s)^2}{\pi n_I Z_I^2 e^4 E^{-3/2}} \left[\log\left(1 + \frac{1}{x}\right) - \frac{1}{1+x}\right]^{-1} \qquad (3)$$

Here, $x = \frac{\hbar^2 q_0^2}{8m^*E}$ with $q_0$ is the Debye screening wave vector expressed as $q_0 = \sqrt{\frac{e^2 n_I}{\epsilon_0 \epsilon K_B T}}$. While $\epsilon_s$, $\epsilon_0$, $n_I$, and $Z_I$ is denoted as the relative dielectric constant, vacuum permittivity, ionized impurity concentration, and impurity charge respectively. The $\tau_{Pol}$ is obtained as:

$$\tau_{pol} = \left[\sum \frac{C(T,E,e_i^{LO}) - A(T,E,e_i^{LO}) - B(T,E,e_i^{LO})}{Z(T,E,e_i^{LO})E^{3/2}}\right]^{-1} \qquad (4)$$

Where, $e_i^{LO}$ represents the longitudinal-optical phonons expressed as $e_i^{LO} = \hbar\omega_{LO}$, in which longitudinal optical angular frequency is denoted as $\omega_{LO}$. More details about the functions A, B, C, and Z can be found elsewhere[79, 81]. The parameters used in these $\tau$ calculaiton such as effective mass (m*), elastic modulus, deformation potential, dominated LO-phonon frequency ($\hbar\omega_{LO}$), high-frequency ($\varepsilon_\infty$) and lattice ($\varepsilon_L$) dielectric constants were summarized in Table S1. Note that the $10^{21}$ cm$^{-3}$ doping concentration is considered for the estimation of the impurity scattering rate (equation 3) for the estimation of total relaxation time $\tau_{tot}$ (equation 1).

The obtained $\tau_{Acp}$, $\tau_{Imp}$, and $\tau_{Pol}$ for p- and n-type carriers as the function of temperature along different transport directions are compared in Fig S6. The $\tau_{Pol}$ is an order of magnitude lower than $\tau_{Acp}$, and $\tau_{Imp}$ for all three considered Janus β-PdXY (X/Y =S, Se, Te) monolayers. Further the $\tau_{tot}$ is also plotted for n and p-type Janus β-PdXY (X/Y =S, Se, Te) monolayers along both x- and y-directions as shown in Fig 7. At room temperature for n- and p-type PdSeTe and PdSTe monolayer, the $\tau_{tot}$ is in the same range along both x and y directions which can be attributed to their nearly equivalent average effective masses. While p-type PdSeS monolayer has relatively smaller relaxation time which is attributed to its larger effective mass in the p-type system. Notably, as temperature increases, the relaxation time decreases, which is consistent with the previous studies.[82] Interestingly, the relatively large anisotropic behavior is observed in the n-type compared to p-type PdSeTe and PdSTe monolayers.

### 3.4 Electronic Transport Properties

Fig. 8(a,b) shows the Seebeck coefficient (S) of Janus β-PdXY (X/Y=S, Se, Te) monolayers at different carrier concentrations and 300-800 K temperature range. For the case of both p-type (Fig. 8(a)) and n-type (Fig. S7(a)) doping, the absolute S decreases as the carrier concentration increases along both x and y-direction. For p-type, the obtained S (hole carrier concentration, N$_h$) corresponding to the maximization of ZT for Janus PdSSe, PdSeTe and PdSTe monolayers is 226 μV/K ($2.90 \times 10^{13}\ cm^{-2}$), 240 μV/K ($2.76 \times 10^{13}\ cm^{-2}$) and 263 μV/K ($2.88 \times 10^{13}\ cm^{-2}$) along the y-direction, respectively. In contrast, the S value corresponding to the maximization of ZT for n-type is lower than p-type along both x and y direction for β-PdXY (X/Y=S, Se, Te) monolayers, whereas for PdSTe alone x-direction, maximum S is 220 μV/K. However, the increasing trend of S with temperature for both p and n-type are consistent with $S \propto T$ relationship

(Fig. 8(b) and S7(b)). Notably, for the PdSeTe monolayer along the y-direction, the S increases to 375 μV/K (at 800 K). These high Seebeck coefficients of Janus PdSeTe is higher than pentagonal PtS$_2$ (227.51 μV/K)[44], PtSe$_2$ (297.40 μV/K)[44], PtTe$_2$ (342.91 μV/K)[44], WS$_2$ (328.15 μV/K)[42], WSSe (322.26 μV/K)[42] and WSTe (322.15 μV/K)[42] monolayers at 300 K. On comparing the S of p-type and n-type doping of these Janus monolayers, it is found that high Seebeck coefficient of p-type doping leads to efficient TE performance than that of n-type doping.

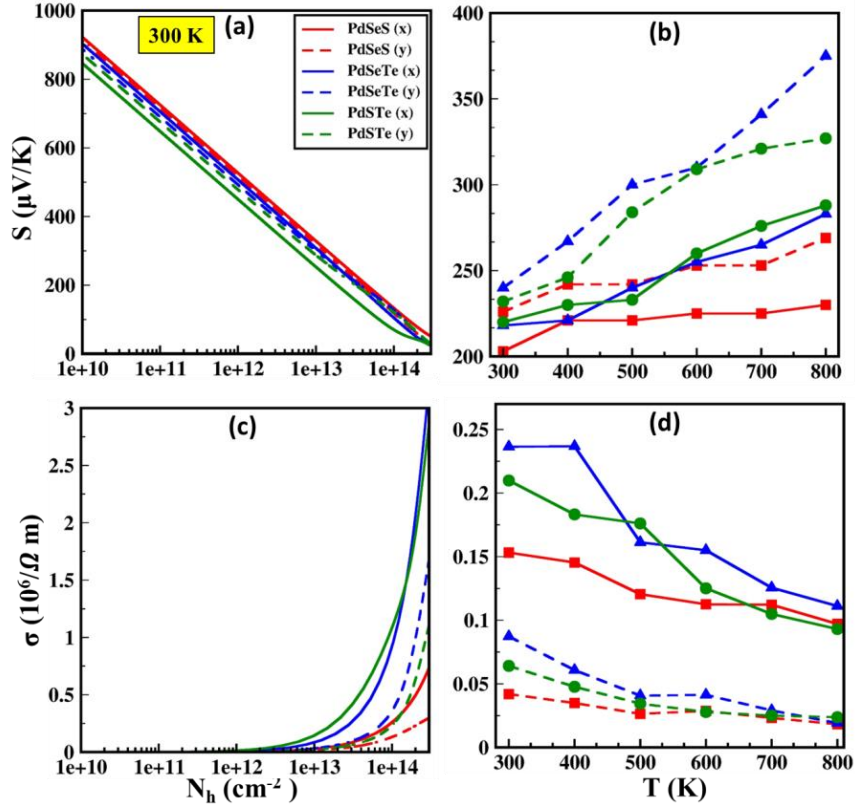

**Fig. 8** The computed Seebeck coefficient (S) and electrical conductivity (σ) as a function of (a, c) carrier concentration ($N_h$) at 300 K and (b, d) temperature (T) for p-type Janus β-PdXY (X/Y =S, Se, Te) monolayers along x and y-direction.

Next, the BTE is used to compute the ratio of electrical conductivity to relaxation time (σ/τ) from the electronic structure of these Janus monolayers. Opposite to S, the carrier concentration increases as σ/τ increases. For Janus β-PdXY (X, Y=S, Se, Te) monolayers, the σ/τ of p-type (Fig. S8 (a)) is significantly higher than that for n-type doping (Fig. S8 (b)). Further, we estimate the electrical conductivity (σ) by substituting the computed total electronic relaxation time ($\tau_{tot}$) into the BTE (Fig. 8(c)). Similar to (σ/τ), the σ of p-type doping is higher than that of their n-type doping. From Fig. 8(d), we demonstrate that the temperature influence on the conductivity is small and consistent with the decreasing trend of S. At 300 K, the highest value of σ is for the Janus PdSTe along the x and y direction in contrast to the highest σ/τ value for the PdSeTe. This dissimilar trend for σ/τ and σ indicates that these Janus monolayers have a significant effect of relaxation time. Notably the larger carrier relaxation time of p-type doping leads to the higher p-type conductivity for all Janus monolayer systems.

Further, we analyze the total thermal conductivity (K), which is defined as the sum of the lattice thermal conductivity ($K_l$) and electronic thermal conductivity ($K_e$), where the $K_e$ is taken from BoltzTraP code.[61] From Fig. 9(a), the p-type Janus PdSeS and PdSeTe monolayer has nearly the same total K along y-direction corresponding to their nearly equivalent $K_e$ value at 300 K. Notably, the small value of total K leads to a high value of ZT. Moreover, the K slowly increases (decreases) with increase in carrier concentration (temperature) for all these Janus monolayers (Fig. 9(b)). Also, similar to $K_e$ the total thermal conductivity (K) of the p-type is more significant than that of the n-type for all three Janus monolayers (Fig. S9(a,b)).

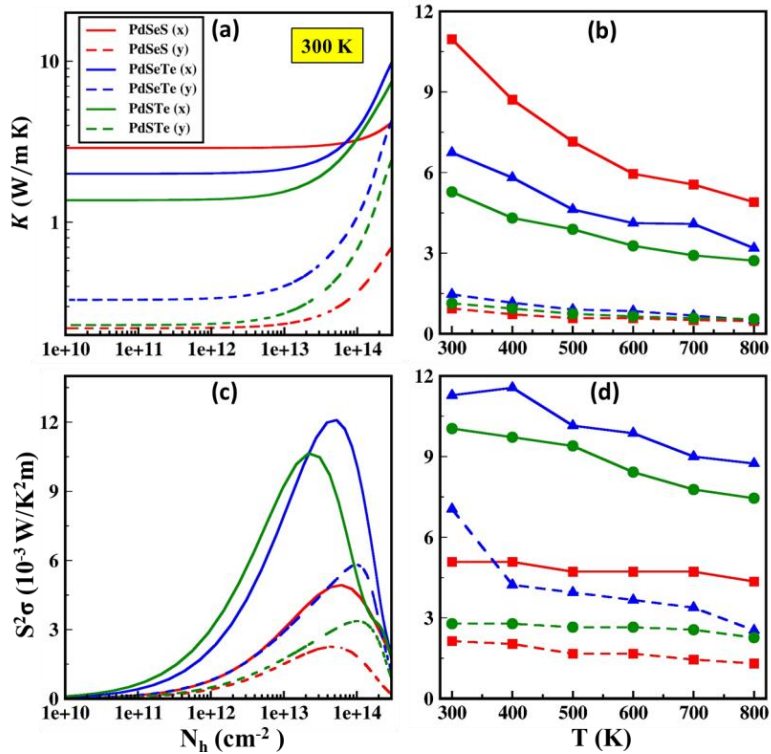

**Fig. 9** The calculated total conductivity (K) and power factor ($S^2σ$) as a function of (a, c) carrier concentration ($N_h$) at 300 K and (b, d) temperature (T) for p-type Janus β-PdXY (X/Y =S, Se, Te) monolayers along x and y-direction.

We further estimate the power factor (PF) using the expression PF= $S^2\sigma$ for the Janus β-PdXY (X/Y=S, Se, Te) monolayers for p-type (Fig. 9 (c,d)) and n-type (Fig. S9(cd)) doping as a function of carrier concentration and temperature along both x and y-direction. AT 300 K, the maximum calculated PF corresponding to the maximization of ZT for p-type are 0.004 (0.002) W/K$^2$m, 0.011 (0.007) W/K$^2$m and 0.010 (0.002) W/K$^2$m along x (y) direction for Janus PdSSe, PdSeTe and PdSTe monolayers, respectively. Moreover, the PF decreases as the temperature increases for all three Janus monolayers corresponding to their value of S and σ (Fig. 9(d)).

Notably, the PF values of the p-type Janus PdSeTe monolayer along x-direction are greatly affected by temperature corresponding to their σ value. At 300 K temperature, the PF values for all these Janus materials with n-type doping (Fig. S8(c,d)) are lower than those of the p-type doping. These calculated maximum PF value at 300 K is less than GeAsSe (0.76 W/K$^2$m)[82], SnSbTe (1.16 W/K$^2$m), InSe (0.049 W/K$^2$m)[83] SnSe (0.020 W/K$^2$m)[84] monolayers and comparable to many potential TE 2D materials, like, Mg$_3$Sb$_2$ (0.004 W/K$^2$m)[85], BiOBr (0.012 W/K$^2$m)[86], Tl$_2$O (0.014 W/K$^2$m)[87], GeAs$_2$ (0.014 W/K$^2$m)[37].

### 3.5 Thermoelectric Figure of Merit

The high value of ZT ($= \frac{S^2\sigma T}{K}$) for the material leads to their high TE performance. Thus, we combined electrons and phonons' transport properties to evaluate the dimensionless quantity ZT for Janus β-PdXY (X/Y=S, Se, Te) monolayers. Notably, the optimal value of PF and thermal conductivity leads to a high ZT value. From Fig. 10(a,c), as the carrier concentration increases, similar to PF, the ZT first increases and then decreases due to the opposite trend of S and the σ. The maximum ZT values for p-type at 300 K are 0.13 (0.68), 0.50 (0.86) and 0.57 (0.68) at specific carrier concentrations of $4.32 \times 10^{13}$ $cm^{-2}$ ($2.90 \times 10^{13}$ $cm^{-2}$), $2.76 \times 10^{13}$ $cm^{-2}$ ($2.76 \times 10^{13}$ $cm^{-2}$) and $1.47 \times 10^{13}$ $cm^{-2}$ ($2.88 \times 10^{13}$ $cm^{-2}$) along x (y)-direction for Janus PdSSe, PdSeTe, and PdSTe monolayers, respectively.

The carrier concentration corresponding to highest ZT value for the temperature range 300 K to 800 K along different direction are illustrated in Table S2. The results for n-type doping are plotted in Fig. 10(c,d), where y (0.76), y (0.84), and y (0.71) direction have maximum ZT value for Janus PdSSe, PdSeTe, and PdSTe monolayers, respectively. Notably, the value of ZT along y-direction at higher temperature is significantly larger (> 3 at 800K) as compared to ZT along x-direction (<

2 at 800 K) (Fig. 10(b,d)). The figure of merit (ZT) generally increases as a function of temperature which can be attributed to the decrease in the total thermal conductivity with temperature (Figure 9(b)). The total thermal conductivity along y-direction decreases from ~1.5 W/mK at 300 K to ~0.5 W/mK at 800 K as compared to the decrease from ~5-11 W/mK at 300 K to ~3-6 W/mK at 800 K along x-direction.

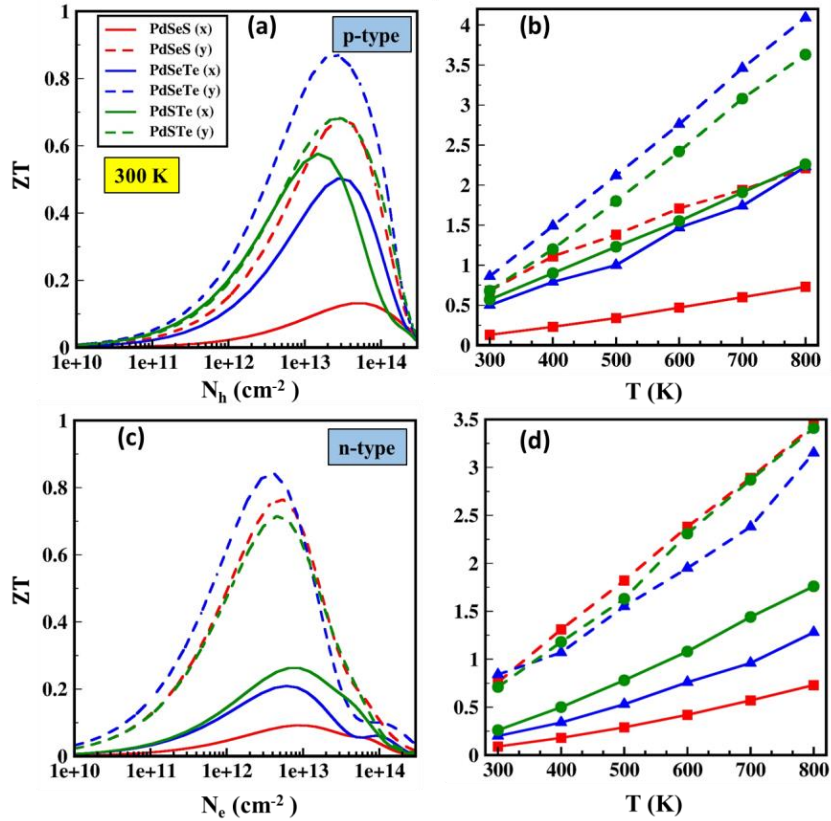

**Fig. 10** The calculated figure of merit (ZT) as a function of (a, c) carrier concentration ($N_h$ and $N_e$) at 300 K and (b, d) temperature (T) for p and n-type Janus β-PdXY (X/Y =S, Se, Te) monolayers along x and y-direction.

Furthermore, the electronic thermal conductivity is defined as $K_e = L\sigma T$ by Wiedemann−Franz law[88], where Lorenz number (L) is useful in a variety of scenarios, such as designing thermoelectric materials and determining the lattice thermal conductivity experimentally. The L of the Wiedemann-Franz law is known to deviate from expected values due to many exotic scenarios such as multi-band and bipolar effect in semiconductors.[89, 90] Thus, here, we also estimate the value of L ($10^{-8}$ WΩK$^{-2}$) separately for all three Janus β-PdXY (X, Y=S, Se, Te) monolayers using the expression: $L = 1.5 + \exp\left[-\frac{|S|}{116}\right]$.[78] The electronic thermal conductivities as functions of carrier concentration at 300 K from both methods are shown in Fig. S10. The calculated revised thermal conductivity ($K_e$) shows minor deviation for both doping types along x and y-direction. We also estimate the effect of revised thermal conductivity ($K_e$) on ZT at 300 K for both p and n-type doping as shown in Fig. S11 (a,b). The change in ZT is more at lower concentration as compared to higher concentration due to the similar trends in the value of $K_e$. ZT value of p-type PdSeTe (PdSTe) monolayer is increased from 0.86 to 0.94 (0.68 to 0.73) along y direction, respectively.

**Comparison with Parent monolayers**

With all of the transport coefficients obtained, we now evaluate the thermoelectric performance of the parent β-PdX$_2$ (X=S, Se, Te) monolayers for the sake of comparison. The lattice thermal conductivity ($K_l$) of β-PdX$_2$ (X=S, Se, Te) monolayer with respect to temperature are summarized in Fig. S12 and compared in Table S3 along x and y-direction. From the calculations, at room temperature the obtained lattice thermal conductivity is 40.58 (7.13), 4.42 (1.47) and 6.80 (3.11) W/m K for β-PdX$_2$ (X=S, Se, Te) monolayers along the x (y) directions, respectively, which are much higher than their considered Janus β-PdXY (X/Y=S, Se, Te) monolayers which may be due to differences in group velocity, anharmonicity and scattering phase space along the different directions. Moreover, among β-phase parent monolayers, the highest $K_l$ of β-PdS$_2$ monolayer can be attributed from its high frequency phonon modes (13.3 THz) as compared to other parent monolayers (< 6 THz).

The $K_l$ of these β-phase Janus monolayers are comparable to the hexagonal phase Janus monolayers along x direction (Table S3), whereas ultra-low $K_l$ value as low as 0.77 W/m K along y-direction is achieved due to the anisotropic structural arrangements of atoms in these monolayers. The estimated ZT value of β Janus (PdSSe and PdSeTe) monolayers are relatively larger than their hexagonal Janus monolayer suggesting these monolayers to be highly efficient TE materials for medium-temperature applications.

## Conclusions

In summary, we systematically explore the TE performance of the 2D Janus β-PdXY (X/Y =S, Se, Te) materials using first principles computation by utilizing the BTE for electrons and phonons. Both dynamic and thermal stability is confirmed by their phonon spectra and AIMD computation for all three Janus materials. At the HSE06 level of theory, the electronic structure computation shows indirect band gap semiconductor of 2.04 eV, 1.67 eV, and 1.71 eV for the Janus PdSSe, PdSeTe and PdSTe monolayers, respectively. The bands of all three Janus materials near the VBM are significantly degenerate. Because of the anisotropic structure of the Janus β-PdXY (X/Y =S, Se, Te) monolayers, the transport properties are highly anisotropic for n and p-type doping along with different directions. Meanwhile, at 300 K, the small phonon group velocity, low converged scattering rate, and high phase space leads to lower $K_l$ of 0.80 W/m K, 0.94 W/m K, and 0.77 W/m K along the y-direction for these Janus materials. From the electronic part, at 500 K, the Seebeck coefficients of these materials are also highly anisotropic and greater than 300 µV/K along both x- y-direction. Combining the low thermal conductivity and high-power factor, the p-type doping has better TE performance comparable to the n-type with as a function of temperature (300 K to 800 K). Meantime, the optimal ZT at room temperature of 0.68 (2.21), 0.86 (4.09) and 0.68 (3.63) is obtained along the y-direction for Janus β-PdXY monolayers. These ultra-high ZT values indicate that the Janus β-PdXY monolayers will exhibit high TE performance for TE applications.

## Supporting Information

AIMD at 300K and 500 K, electronic properties at PBE+SOC level of theories, PDOS, electronic transport, and thermoelectric ZT for the n-type Janus β-PdXY (X/Y=S,Se,Te) monolayers.

## Acknowledgements

The computational facility available at the Himachal Pradesh University, Shimla and Central University of Punjab, Bathinda, was used to obtain the results presented in this paper. MJ is thankful to the University Grants Commission (UGC) for financial assistance in the form of the Senior Research Fellowship. Helpful discussion with Jaspreet Singh and Poonam is highly acknowledged.